\documentclass[preprint,aps,jmp,showpacs,nofootinbib]{revtex4}

\usepackage{amsbsy}
\usepackage{mathrsfs}

\begin{document}

\title{Covariant canonical formalism for four-dimensional BF theory}

\date{\today}

\author{Mauricio Mondrag\'on\footnote{Current address: Centre de Physique
Th\'eorique, Campus de Luminy, F-13288 Marseille, France. Electronic address:
mmondrag@cpt.univ-mrs.fr }} \email{mo@fis.cinvestav.mx}
\affiliation{Departamento de F\'{\i}sica, Cinvestav, Av. Instituto
Polit\'ecnico Nacional 2508, San Pedro Zacatenco, 07360, Gustavo A. Madero,
Ciudad de M\'exico, M\'exico}

\author{Merced Montesinos\footnote{Associate Member of the Abdus Salam
International Centre for Theoretical Physics, Trieste, Italy.}}
\email{merced@fis.cinvestav.mx}
\affiliation{Departamento de F\'{\i}sica,
Cinvestav, Av. Instituto Polit\'ecnico Nacional 2508, San Pedro Zacatenco,
07360, Gustavo A. Madero, Ciudad de M\'exico, M\'exico}

\begin{abstract}
The covariant canonical formalism for four-dimensional BF theory is performed.
The aim of the paper is to understand in the context of the covariant
canonical formalism both the reducibility that some first class constraints
have in Dirac's canonical analysis and also the role that topological terms
play. The analysis includes also the cases when both a cosmological constant
and the second Chern character are added to the pure BF action. In the case of
the BF theory supplemented with the second Chern character, the presymplectic
3-form is different to the one of the BF theory in spite of the fact both
theories have the same equations of motion while on the space of solutions
they both agree to each other. Moreover, the analysis of the degenerate
directions shows some differences between diffeomorphisms and internal gauge
symmetries.
\end{abstract}


\maketitle

\section{Introduction}
In the path integral quantization of a given field theory one needs to sum the
exponential of the classical action weighted with a suitable factor over all
possible configurations of the fields under consideration. If the theory under
study is a gauge theory one needs, in addition, to factor out the gauge
transformations in such a way that the sum includes only equivalence classes
of gauge transformed fields. So, intuitively, it is expected that any change
in the action principle, yielding the same classical equations of motion,
provides a completely different quantum theory. For instance, if the
Yang-Mills Lagrangian density $\mbox{tr} F \wedge \star F$ is modified adding
the term $\theta \mbox{tr} F \wedge F$ the resulting quantum theory is
sensitive to this contribution even when it does not modify the classical
equations of motion \cite{Jackiw}. One way to understand, at the classical
level, the cause of having a different quantum theory for the Yang-Mills field
is to realize that, in the generic case, the specification of the Lagrangian
density is equivalent to specify the symplectic geometry in the various phase
spaces associated with the classical theory. Thus, if the Lagrangian density
changes, the symplectic geometry also does generically. If one accepts that
what defines a dynamical system is its equations of motion then this knowledge
is not enough to specify the symplectic geometry on the various phase spaces
involved. If, one the other hand, one accepts that what defines a dynamical
system is its equations of motion plus an action principle (which provides its
equations of motion), then one is in a different situation. The difference is
that, as we have already mentioned, the specification of the Lagrangian
density provides a symplectic structure. Thus, an action principle plays a
double role: (1) it provides the equations of motion and also (2) provides the
symplectic geometry. Before going into the analysis of the BF theory, which is
the subject of this paper, let us emphasize this point with a very simple
example borrowed from dynamical systems with a finite number of degrees of
freedom. The equations of motion for the two-dimensional isotropic harmonic
oscillator ${\ddot x} + \omega^2 x =0$ and ${\ddot y} +\omega^2 y=0$ can be
obtained from the Lagrangian ${\cal L}(x,y,{\dot x},{\dot y})=m ({\dot x}
{\dot y} - \omega^2 xy)$ or from ${\cal L}_{usual} (x,y,{\dot x},{\dot y})=
\frac12 m \left ( {\dot x}^2 + {\dot y}^2 - \omega^2 x^2 - \omega^2 y^2 \right
)$. Moreover, ${\cal L} \neq {\cal L}_{usual} + d F(x,y,t)/dt$. Note also that
we are not making a change of coordinates, which are the same for both cases.
The symplectic structures coming from these Lagrangians are very distinct to
each other even when they both provide the same equations of motion
\cite{gerardo}. Coming back to field theory, it has been shown, in the context
of Dirac's canonical analysis, that the symplectic potential changes if one
adds topological terms to the Lagrangian density of tetrad gravity
\cite{MonCQGQ01}. On the other hand, using the covariant canonical formalism,
it has been shown that the inclusion of topological terms in Lagrangians for
string theory also modifies the original symplectic potential
\cite{Cartas,Cartas2}.

In this paper, in the context of the covariant canonical formalism
\cite{Witten,Crn,Lee,ABR}, we perform the covariant canonical analysis of
four-dimensional BF theory, BF theory plus a cosmological constant $\Lambda$,
and BF plus the second Chern character $F^{IJ} \wedge F_{IJ}$. These theories
are topological in the sense that there is no fixed background metric $g$ on
the four-dimensional manifold ${\cal M}$ in which they are defined. In
addition, they are topological in the sense that they have no local degrees of
freedom. A more detailed analysis of the covariant canonical formalism for BF
theory can be found in Ref. \cite{MO}. Of course, the inclusion of the second
Chern character does not modify the equations of motion while the cosmological
constant does. However, the aim of the paper is to study the symplectic
geometry involved. It must be emphasized that Dirac's canonical analysis for
BF theory has been already done \cite{Gary,Caicedo} (see also the Appendix A).
In Ref. \cite{Caicedo} it is shown that the first class constraints
${\widetilde \Psi}^a\,_{IJ}:=  \frac12 {\widetilde \eta}^{abc} F_{bcIJ} (A)
\approx 0$ are reducible. Dirac's canonical analysis for BF theory with a
cosmological constant is reported in the Appendix A, where it is shown that
now the reducibility equations involve both the Gauss constraints $
{\widetilde \Psi}^{IJ}$ and the other set of first class constraints
${\widetilde \Psi}^a\,_{IJ} :=  \frac12 {\widetilde \eta}^{abc} F_{bcIJ} -
\Lambda \varepsilon_{IJKL} {\widetilde \Pi}^{aKL} \approx 0$. This is so
because of the cosmological constant $\Lambda$. In both cases, the
reducibility equations in Dirac's canonical analysis come from the Bianchi
identities $DF_{IJ}=0$. So, the covariant canonical formalism is an
opportunity to understand the role these identities play on this formalism.

\section{BF theory}
The four-dimensional BF theory with $SO(3,1)$ as the internal relevant group
is defined by the equations of motion
\begin{eqnarray}\label{em}
F_{IJ}=0 \, , \quad D B^{IJ}= 0 \, ,
\end{eqnarray}
where $F_{IJ}(A)= d A_{IJ} + A_{IK} \wedge A^K\,_J$ is the curvature of the
Lorentz connection 1-form $A_{IJ}$, $B^{IJ}=\frac12 B^{IJ}_{\alpha\beta} d
x^{\alpha} \wedge d x^{\beta}$ is a set of six 2-forms, $D B^{IJ} := d B^{IJ}
+ A^I\,_K \wedge B^{KJ} + A^J\,_K \wedge B^{IK}$ is the covariant derivative
of $B^{IJ}$; $I,J=0,1,2,3$ are Lorentz indexes which are raised and lowered
with the Minkowski metric $\eta_{IJ}$. Even though the analysis will be
restricted to a Lorentz BF theory, the results are generic in the sense that
hold for any BF theory in 4-spacetime dimensions. The choice of the Lorentz
group is only to fix the notation that might be used for the case of BF
gravity.

In the context of the covariant canonical formalism, the {\it kinematical}
phase space ${\mathscr{F}}$ of the theory is defined as the space formed by
all smooth Lorentz connections $A_{IJ}$ and $B^{IJ}$ fields. Any generic point
of ${\mathscr{F}}$ is not required to satisfy the equations of motion of the
BF theory. The space of solutions to the equations of motion
$\overline{\mathscr{F}}$ is considered as submanifold of ${\mathscr{F}}$ and
is formed by all points of ${\mathscr{F}}$ that satisfy the equations of
motion of Eq. (\ref{em}). The reduced (or physical) phase space for the theory
is reached by making the quotient of $\overline{\mathscr{F}}$ by the gauge
transformations of the theory \cite{Witten,Crn}. Even though the term ``phase
space" has been used to name these different manifolds, it must be emphasized
that, at this stage, ${\mathscr{F}}$, $\overline{\mathscr{F}}$, and the
reduced phase space carry no intrinsic symplectic geometry. Thus, although the
equations of motion of Eq. (\ref{em}) are used to define
$\overline{\mathscr{F}}$, they are not enough to uniquely endow the various
phase spaces for the theory already mentioned with symplectic geometry. Where
does symplectic geometry come from then? One possibility is from action
principles, specifying the Lagrangian density \cite{mau}. The equations of
motion of Eq. (\ref{em}) are usually obtained from the action \cite{Gary}
\begin{eqnarray}
S[A, B] & = & \int_{{\cal M}} B^{IJ} \wedge F_{IJ} [A] \, . \label{BFaction}
\end{eqnarray}
[see also Refs. \cite{solo} and \cite{me} for alternative choices of the
action]. To get the geometry, one needs to proceed along the following lines.
The first order variation of the Lagrangian 4-form ${\bf L}[A,B]= B^{IJ}
\wedge F_{IJ} [A]$ is
\begin{eqnarray}
\delta {\bf L}[A,B] & = & (\delta B^{IJ}) \wedge F_{IJ} - \left ( D B^{IJ}
\right ) \wedge \delta A_{IJ} + d {\bf \Theta} (B, \delta A) \, ,
\label{VarBFLag}
\end{eqnarray}
from which the {\it presymplectic potential} $3$-form
\begin{eqnarray}
{\bf \Theta} (B,\delta A) := B^{IJ} \wedge \delta A_{IJ} \, , \label{PrePotBF}
\end{eqnarray}
is read off. Now, by taking into account an arbitrary smooth two-parameter
family of field configurations and computing the antisymmetric combination of
the variations in ${\bf L}[A,B]$, $[\delta_1 , \delta_2] {\bf L}[A,B]=0$
yields \cite{Lee}
\begin{eqnarray}
d \omega (\delta_1 A , \delta _1 B , \delta_2 A , \delta_2 B) & = & (\delta_1
B^{IJ}) \wedge \delta_2 F_{IJ} - \left ( \delta_2 D B^{IJ} \right ) \wedge
\delta_1 A_{IJ} - \left ( \delta_1 \longleftrightarrow \delta_2 \right ) \, ,
\end{eqnarray}
where
\begin{eqnarray}
\omega (\delta_1 A , \delta _1 B , \delta_2 A , \delta_2 B) & = &  \left (
\delta_1 B^{IJ} \wedge \delta_2 A_{IJ} - \delta_2 B^{IJ} \wedge \delta_1
A_{IJ} \right ) \label{PreStru}\, .
\end{eqnarray}
is the {\it presymplectic} $3$-form \cite{Baez}.

\subsection{``Fundamental'' set of local gauge transformations}\label{FundSet}
(i) {\it Local Lorentz transformations}: The action is fully gauge invariant
under any arbitrary finite local Lorentz transformation. The infinitesimal
version of this transformation is
\begin{eqnarray}
\delta_{\varepsilon} A_{IJ} & = & D \varepsilon_{IJ} \, , \nonumber\\
\delta_{\varepsilon} B^{IJ} & = & - \varepsilon^I\,_K B^{KJ} -
\varepsilon^J\,_K B^{IK} \, , \label{IntSymm}
\end{eqnarray}
where $\varepsilon_{IJ}$ are the infinitesimal gauge parameters. The change of
the Lagrangian ${\bf L}[A,B]$ induced by the infinitesimal variation of the
fields, given in Eq. (\ref{IntSymm}), is
\begin{eqnarray}\label{VarLor}
\delta_{\varepsilon} {\bf L} [A,B] & = &  \delta_{\varepsilon} B^{IJ} \wedge
F_{IJ} +  B^{IJ} \wedge \delta_{\varepsilon} F_{IJ} \nonumber\\
& = & 0 \, .
\end{eqnarray}
Therefore, from Eqs. (\ref{VarBFLag}) and (\ref{VarLor}) the {\it Noether
current} 3-form ${\bf J}_N [A,B,\varepsilon]$ (Ref. \cite {Lee}) associated
with the symmetry (\ref{IntSymm}) is \cite{MonCQG03}
\begin{eqnarray}
{\bf J}_N [A,B,\varepsilon] & = & {\bf \Theta} (B, D \varepsilon_{IJ})
\nonumber\\
& = &  B^{IJ} \wedge D \varepsilon_{IJ} \, ,
\end{eqnarray}
which can be rewritten as
\begin{eqnarray}
{\bf J}_N [A,B,\varepsilon] & = & d {\bf Q}[B,\varepsilon] - \varepsilon_{IJ}
\wedge D B^{IJ} \, , \label{NoetherC}
\end{eqnarray}
with
\begin{eqnarray}
{\bf Q}[B, \varepsilon] := \varepsilon_{IJ} B^{IJ} \, ,
\end{eqnarray}
the corresponding {\it Noether current potential} $2$-form. Equation
(\ref{NoetherC}) has the {\it same} structure that appears in the Noether
current associated with infinitesimal diffeomorphisms in theories with
dynamical background metric in the sense that the right-hand side of Eq.
(\ref{NoetherC}) is the exterior derivative of the Noether current potential
2-form ${\bf Q}[B, \varepsilon]$ plus a term proportional to (one set of) the
equations of motion \cite{Iyer95}. There is {\it a priori} no reason for an
internal symmetry, like (\ref{IntSymm}), behaves in the same manner as
diffeomorphisms.

{\it Degenerate directions}: These can be obtained from the symplectic inner
product between the gauge transformation $\delta_{\varepsilon}$ and an
arbitrary variation $\delta$ by taking $\delta_1 \equiv \delta$ and $\delta_2
=\delta_{\varepsilon}$. From Eqs. (\ref{PreStru}) and (\ref{IntSymm}),
\begin{eqnarray}
\omega(\delta A, \delta B , \delta_{\varepsilon} A , \delta_{\varepsilon} B) &
= & d \left (  \varepsilon_{IJ} \delta B^{IJ} \right ) - \varepsilon_{IJ}
\delta (D
B^{IJ}) \nonumber\\
& = & d (\delta {\bf Q} [B,\varepsilon]) -  \varepsilon_{IJ} \delta (D B^{IJ})
\, . \label{Gauomega}
\end{eqnarray}
Note that on the right-hand side of Eq. (\ref{Gauomega}) appears one term
involving the linearized Eulerian derivative, $\delta (D B^{IJ})$, but no
terms proportional to Eulerian derivatives themselves appear explicitly. As it
will be seen, later on, this is a difference with respect to infinitesimal
diffeomorphisms (see Sec. \ref{candiff}). Thus, we have
\begin{eqnarray}\label{sipI}
\Omega_{\Sigma} (\delta A, \delta B , \delta_{\varepsilon} A ,
\delta_{\varepsilon} B) & := & \int_{\Sigma} \omega(\delta A, \delta B ,
\delta_{\varepsilon} A , \delta_{\varepsilon} B) \nonumber\\
& = & -  \int_{\Sigma} \varepsilon_{IJ} \delta D B^{IJ} + \int_{\partial
\Sigma} (  \varepsilon_{IJ} \delta B^{IJ}) \, .
\end{eqnarray}
The integral over $\Sigma$ depends on the gauge parameters $\varepsilon_{IJ}$,
the fields $(A_{IJ},B^{KL})$ and their variations while the integral over
$\partial\Sigma$ depends only on the gauge parameters $\varepsilon_{IJ}$ and
the variation of the $B^{IJ}$ fields, $\delta B^{IJ}$. Both integrals, in
general, do not vanish and therefore the gauge transformation of Eq.
(\ref{IntSymm}) does not qualify as a degenerate direction unless additional
assumptions are imposed. In particular, one has the following.

{\it Proposition}: If the linearized Eulerian derivative $\delta (D B^{IJ})$
vanishes, $\delta (D B^{IJ})=0$, and the arbitrary variations $\delta B^ {IJ}$
have compact support in the interior of $\Sigma$, $\delta B^{IJ}
\mid_{\partial\Sigma}=0$, then
\begin{eqnarray}\label{kernelI}
\Omega_{\Sigma} (\delta A,\delta B, \delta_{\varepsilon} A,
\delta_{\varepsilon} B) & = & 0 \, ,
\end{eqnarray}
without imposing any additional restrictions on the gauge parameters
$\varepsilon_{IJ}$. Note that $(A_{IJ},B^{KL})$ need not be a point of the
space of solutions to the equations of motion $\overline{\mathscr{F}}$ in
order for Eq. (\ref{kernelI}) to hold (see also Ref. \cite{gregory}).
Nevertheless, it is a common fact to restrict the analysis to
$\overline{\mathscr{F}}$ and also to take $(\delta A_{IJ}, \delta B^{KL})$ as
tangent vectors to $\overline{\mathscr{F}}$. Of course, the integral over
$\partial \Sigma$ in Eq. (\ref{sipI}) also vanishes if the gauge parameters
$\varepsilon_{IJ}$ vanish at $\partial\Sigma$, i.e., if the infinitesimal
gauge transformation of Eq. (\ref{IntSymm}) is the identity at $\partial
\Sigma$.

{\it Canonical transformations}:

{\it Proposition}: The infinitesimal gauge transformation of Eq.
(\ref{IntSymm}) is a canonical transformation.

{\it Proof}: From the gauge transformation of Eq. (\ref{IntSymm}),
\begin{eqnarray}\label{GaugeTF}
{A'}_{IJ} & = & A_{IJ} + D \varepsilon_{IJ} \, , \nonumber\\
{B'}^{IJ} & = & B^{IJ} - \varepsilon^I\,_K B^{KJ} - \varepsilon^J\,_K B^{IK}
\, ,
\end{eqnarray}
we can compute two arbitrary variations of the gauge-transformed fields of Eq.
(\ref{GaugeTF}) (Ref. \cite{Crn}),
\begin{eqnarray}
\delta_i {A'}_{IJ} & = & \delta_i A_{IJ} - \delta_i A^K\,_I \,\,
\varepsilon_{KJ}
- \delta_i A^K\,_J \,\, \varepsilon_{IK} \, , \nonumber\\
\delta_i {B'}^{IJ} & = & \delta_i B^{IJ} - \varepsilon^I\,_K  \delta_i B^{KJ}
- \varepsilon^J\,_K \delta_i B^{IK} \, .
\end{eqnarray}
So,
\begin{eqnarray}
{\omega'} & := & \left ( \delta_1 {B'}^{IJ} \wedge \delta_2 {A'}_{IJ} -
\delta_2
{B'}^{IJ} \wedge \delta_1 {A'}_{IJ} \right ) \nonumber\\
& = &  \left (\delta_1 B^{IJ} - \varepsilon^I\,_K  \delta_1 B^{KJ} -
\varepsilon^J\,_K \delta_1 B^{IK}\right ) \wedge \left ( \delta_2 A_{IJ} -
\delta_2 A^K\,_I \,\, \varepsilon_{KJ} - \delta_2 A^K\,_J \,\,
\varepsilon_{IK}\right ) \nonumber\\
& & -  \left (\delta_2 B^{IJ} - \varepsilon^I\,_K  \delta_2 B^{KJ} -
\varepsilon^J\,_K \delta_2 B^{IK}\right ) \wedge \left ( \delta_1 A_{IJ} -
\delta_1 A^K\,_I \,\, \varepsilon_{KJ} - \delta_1 A^K\,_J \,\,
\varepsilon_{IK}\right ) \nonumber\\
& = & \omega \, ,
\end{eqnarray}
exactly, i.e., without using any additional conditions. Therefore,
\begin{eqnarray}
{\Omega'}_{\Sigma} & : = & \int_{\Sigma} \omega (\delta_1 {A'}, \delta_1 {B'},
\delta_2 {A'}, \delta_2 {B'} ) \nonumber\\
& = & \int_{\Sigma} \omega (\delta_1 A , \delta_1 B, \delta_2 A , \delta_2 B )
= \Omega_{\Sigma} \, .
\end{eqnarray}

(ii) {\it $B$'s transform like connections}: The infinitesimal version of this
gauge transformation is
\begin{eqnarray}
\delta_{\chi} A_{IJ} & = & 0 \, , \nonumber\\
\delta_{\chi} B^{IJ} & = & D \chi^{IJ} \, , \label{ExtSym}
\end{eqnarray}
where the gauge parameters $\chi^{IJ}$ are $1$-forms. However, this symmetry
is peculiar in the sense that it does not satisfy the definition of symmetry
in a strict sense \cite{Lee}. To see this, the variation of the Lagrangian
${\bf L}[A,B]$ induced by the variation of the fields is computed
\begin{eqnarray}\label{2ndGT}
\delta_{\chi} {\bf L}[A,B] & = &  D \chi^{IJ} \wedge F_{IJ} \, ,
\end{eqnarray}
which has not the desired form in the sense that the right-hand side of Eq.
(\ref{2ndGT}) is not of the form $d \boldsymbol{\alpha}$. To continue, we must
rewrite the right-hand side of the last equation,
\begin{eqnarray}
\delta_{\chi} {\bf L}[A,B] & = & d \left (  \chi^{IJ} \wedge F_{IJ} \right ) +
\chi^{IJ} \wedge D F_{IJ} \, . \label{ChiVar}
\end{eqnarray}
Thus, the right-hand side of Eq. (\ref{ChiVar}) is not, in a strict sense, of
the form $d \boldsymbol{\alpha}$. It acquires this form just if the Bianchi
identities $D F_{IJ}=0$ are used. However, when computing the transformation
of the Lagrangian ${\bf L}[A,B]$ induced by the transformation of the fields
it is not allowed to use the equations of motion in order to check if the
transformation of the fields does (or does not) qualify as a gauge symmetry. A
purist might say that the second term on the right-hand side of Eq.
(\ref{ChiVar}) involves no equations of motion simply because the Bianchi
identities do {\it not} qualify as equations of motion in the sense that they
do not appear when the first order variation of the Lagrangian is computed
[see Eq. (\ref{VarBFLag})].

Therefore, from Eqs. (\ref{VarBFLag}) and (\ref{ChiVar}),
\begin{eqnarray}
d \left (  \chi^{IJ} \wedge F_{IJ} \right ) +  \chi^{IJ} \wedge D F_{IJ}  & =
& (D \chi^{IJ}) \wedge F_{IJ}  \, ,
\end{eqnarray}
and so
\begin{eqnarray}
d {\bf J}_N [A,\chi] & = &  -  \chi^{IJ} \wedge D F_{IJ} +  (D \chi^{IJ})
\wedge F_{IJ} \, ,
\end{eqnarray}
with \cite{MonCQG03}
\begin{eqnarray}
{\bf J}_N [A,\chi] & := &  \chi^{IJ} \wedge F_{IJ} \, ,
\end{eqnarray}
the Noether current associated with the local symmetry (\ref{ExtSym}). Note
that ${\bf J}_N [A,\chi]$ is proportional to the Eulerian derivative $F_{IJ}$.
Note that if the equations of motion hold (i.e., if $F_{IJ}=0$ hold) and the
Bianchi identities hold (i.e, if $D F_{IJ}=0$ hold) then the Noether current
is identically conserved. Moreover, note that ${\bf J}_N [A,\chi]$ identically
vanishes on-shell, i.e., ${\bf J}_N =0$ if $F_{IJ}=0$.

{\it Degenerate directions}: Again, from the symplectic inner product between
the gauge transformation $\delta_{\chi}$ and an arbitrary variation $\delta $
and Eqs. (\ref{PreStru}) and (\ref{ExtSym})
\begin{eqnarray}
\omega(\delta A,\delta B, \delta_{\chi} A, \delta_{\chi} B) & = &  d (\delta
\left ( - \chi^{IJ} \wedge  A_{IJ} \right )) -  \chi^{IJ} \wedge \delta F_{IJ}
\, . \label{chiomega}
\end{eqnarray}
Note that on the right-hand side of Eq. (\ref{chiomega}) appears the
linearized Eulerian derivative $\delta F_{IJ}$ but not the Eulerian
derivatives themselves in contrast to what happens with diffeomorphisms (see
Sec. \ref{candiff}). Thus, we have
\begin{eqnarray}\label{sipII}
\Omega_{\Sigma} (\delta A,\delta B, \delta_{\chi} A, \delta_{\chi} B) & := &
 \int_{\Sigma} \omega(\delta A,\delta B, \delta_{\chi} A,
\delta_{\chi} B) \nonumber\\
& = & -  \int_{\Sigma} \chi^{IJ} \wedge \delta F_{IJ} -  \int_{\partial
\Sigma} ( \chi^{IJ} \wedge  \delta A_{IJ} ) \, .
\end{eqnarray}
Again, the integral over $\Sigma$ depends on the gauge parameters $\chi^{IJ}$,
the field $A_{IJ}$ and its first order variations $\delta A_{IJ}$ while the
integral over $\partial \Sigma$ depends only on the gauge parameters
$\chi^{IJ}$ and the variations of the field $A_{IJ}$, $\delta A_{IJ}$. Both
integrals, in general, do not vanish and therefore the gauge transformation of
Eq. (\ref{ExtSym}) does not qualify as a degenerate direction unless
additional assumptions are imposed. In particular, one has the following.

{\it Proposition}. If the linearized Eulerian derivative $\delta F_{IJ}$
vanishes, $\delta F_{IJ}=0$, and the arbitrary variations $\delta A_{IJ}$ have
compact support in the interior of $\Sigma$, $\delta A_{IJ}
\mid_{\partial\Sigma}=0$, then
\begin{eqnarray}\label{kernelII}
\Omega_{\Sigma} (\delta A,\delta B, \delta_{\chi} A, \delta_{\chi} B) & = & 0
\, ,
\end{eqnarray}
without imposing any additional conditions on the gauge parameters
$\chi^{IJ}$. Note also that in order for Eq. (\ref{kernelII}) to hold it is
not necessary that the point $(A_{IJ},B^{KL})$ belongs to the space of
solutions to the equations of motion $\overline{\mathscr{F}}$. Nevertheless,
it is a common fact to restrict the analysis to this case and also to take
$(\delta A_{IJ}, \delta B^{KL})$ as tangent vectors to
$\overline{\mathscr{F}}$. Of course, the integral over $\partial \Sigma$ in
Eq. (\ref{sipII}) also vanishes if the gauge parameters $\chi^{IJ}$ vanish at
$\partial\Sigma$, i.e., if the infinitesimal gauge transformation of Eq.
(\ref{ExtSym}) is the identity at $\partial\Sigma$.

{\it Canonical transformations}:

{\it Proposition}: The transformation induced by the gauge symmetry of Eq.
(\ref{ExtSym}) is an infinitesimal canonical transformation.

{\it Proof}: In fact, from Eq. (\ref{ExtSym}),
\begin{eqnarray}
{A'}_{IJ} & = & A_{IJ} \, , \nonumber\\
{B'}^{IJ} & = & B^{IJ} + D \chi^{IJ} \, ,
\end{eqnarray}
we can compute two arbitrary variations of the gauge-transformed fields
\cite{Crn}
\begin{eqnarray}
\delta_i {A'}_{IJ} & = & \delta_i A_{IJ} \, , \nonumber\\
\delta_i {B'}^{IJ} & = & \delta_i B^{IJ} + \delta_i A^I\,_K \wedge \chi^{KJ} +
\delta_i A^J\,_K \wedge \chi^{IK} \, , \quad i=1,2 \, .
\end{eqnarray}
So,
\begin{eqnarray}
{\omega'} & := &  \left ( \delta_1 {B'}^{IJ} \wedge \delta_2 {A'}_{IJ} -
\delta_2
{B'}^{IJ} \wedge \delta_1 {A'}_{IJ} \right ) \nonumber\\
& = &  \left ( \delta_1 B^{IJ} + \delta_1 A^I\,_K \wedge \chi^{KJ} + \delta_1
A^J\,_K \wedge \chi^{IK} \right ) \wedge \delta_2 A_{IJ} \nonumber\\
& & -  \left ( \delta_2 B^{IJ} + \delta_2 A^I\,_K \wedge \chi^{KJ} + \delta_2
A^J\,_K \wedge \chi^{IK} \right ) \wedge \delta_1 A_{IJ} \nonumber\\
& = & \omega \, ,
\end{eqnarray}
exactly, i.e., without using any additional conditions. Therefore,
\begin{eqnarray}
{\Omega'}_{\Sigma} & : = & \int_{\Sigma} \omega (\delta_1 {A'}, \delta_1 {B'},
\delta_2 {A'}, \delta_2 {B'})  \nonumber\\
& = & \int_{\Sigma} \omega (\delta_1 A , \delta_1 B, \delta_2 A , \delta_2 B )
= \Omega_{\Sigma} \, ,
\end{eqnarray}
under the infinitesimal gauge transformation of Eq. (\ref{ExtSym}).

\subsection{Diffeomorphisms}\label{candiff}
The gauge symmetries discussed in Sec. \ref{FundSet} can also be obtained by
using Dirac's canonical analysis. In addition, Dirac's canonical analysis
shows that the full set of constraints are first class. There are no second
class constraints in the theory. However, the first class constraints
${\widetilde \Psi}^a\,_{IJ}$ (which generate the $\delta_{\chi}$ symmetry) are
{\it reducible} on account of the Bianchi identities $D F_{IJ}=0$ which imply
the reducibility equation $D_a {\widetilde \Psi}^a\,_{IJ} = 0$. Once
reducibility is taken into account the counting of the local degrees of
freedom is zero, showing that the theory has only global degrees of freedom
(see, for instance, Ref. \cite{Caicedo} and the Appendix A). Moreover, it is
also known that the theory is diffeomorphism covariant. Therefore, the
transformation of the fields induced by diffeomorphisms must be built from the
``fundamental" set of gauge transformations (\ref{IntSymm}) and
(\ref{ExtSym}). (The quotation marks in the word ``fundamental" emphasize the
fact that the gauge transformations are not independent on account of the
reducibility of the constraints.) In fact, a diffeomorphism induces a change
in the fields $A_{IJ}$ given by
\begin{eqnarray}\label{mau}
\delta_{\xi} A_{IJ} & = & {\cal L}_{\xi} A_{IJ} =  \xi \cdot F_{IJ} + D
\varepsilon_{IJ}  =  \xi \cdot F_{IJ} + \delta_{\varepsilon} A_{IJ}\, ,
\end{eqnarray}
as well as in the fields $B^{IJ}$,
\begin{eqnarray}
\delta_{\xi} B^{IJ} & = & {\cal L}_{\xi} B^{IJ}  =  \xi \cdot D B^{IJ} -
\varepsilon^I\,_K B^{KJ} - \varepsilon^J\,_K B^{IK} + D \chi^{IJ} \nonumber\\
& = & \xi \cdot D B^{IJ} + \delta_{\varepsilon} B^{IJ} + \delta_{\chi} B^{IJ}
\, ,
\end{eqnarray}
where $\varepsilon_{IJ}:= \xi \cdot A_{IJ}$ is a set of $0$-forms and
$\chi^{IJ}:= \xi \cdot B^{IJ}$ is a set of $1$-forms.

{\it Noether current}: The {\it Noether current} $3$-form associated with
diffeomorphisms is \cite{MonCQG03}
\begin{eqnarray}
{\bf J}_N [A,B,\xi] & = & {\bf \Theta}(B,{\cal L}_{\xi} A) - \xi \cdot {\bf L}
\nonumber\\
& = & d {\bf Q} [A,B,\xi] -  (\xi \cdot A_{IJ}) D B^{IJ} - (\xi \cdot B^{IJ})
\wedge F_{IJ} \, , \label{VarJ}
\end{eqnarray}
where
\begin{eqnarray}
{\bf Q}[A,B,\xi] = (\xi \cdot A_{IJ}) B^{IJ} \, ,
\end{eqnarray}
is the {\it Noether current potential} $2$-form. If $(A_{IJ}, B^{IJ})$ is a
point of the space of solutions to the equations of motion
$\overline{\mathscr{F}}$ then ${\bf J}_N [A,B,\xi]$ can be obtained from the
Noether current potential ${\bf J}_N [A, B,\xi] = d {\bf Q} [A,B,\xi]$. The
{\it Noether charge} ${\cal Q}_{\Sigma}(\xi)$ associated with infinitesimal
diffeomorphisms is given by the integral of ${\bf J}_N [A,B,\xi]$ over
$\Sigma$,
\begin{eqnarray}
{\cal Q}_{\Sigma} (\xi) & := & \int_{\Sigma} {\bf J}_N [A,B,\xi] \nonumber\\
& = & \int_{\partial \Sigma} {\bf Q} [A,B,\xi] - \int_{\Sigma} \left [  (\xi
\cdot A_{IJ}) D B^{IJ} +  (\xi \cdot B^{IJ}) \wedge F_{IJ} \right ] \, .
\end{eqnarray}

{\it Proposition}: If $(A_{IJ},B^{IJ})$ is a point in ${\overline\mathscr{F}}$
then the Noether charge just has a contribution from the boundary of $\Sigma$,
\begin{eqnarray}
{\cal Q}_{\Sigma} (\xi) & = & \int_{\partial \Sigma} {\bf Q} [A,B,\xi] =
\int_{\partial\Sigma}  (\xi \cdot A_{IJ}) B^{IJ} \, .
\end{eqnarray}

{\it Canonical transformation induced by diffeomorphisms}: The transformation
of the fields induced by a diffeomorphism is
\begin{eqnarray}\label{4diff}
{A'}_{IJ} & = & A_{IJ} + {\cal L}_{\xi} A_{IJ} \, ,\nonumber\\
{B'}^{IJ} & = & B^{IJ}+ {\cal L}_{\xi} B^{IJ} \, ,
\end{eqnarray}
and so \cite{Crn}
\begin{eqnarray}
\delta_i {A'}_{IJ} & = & \delta_i A_{IJ} + \delta_i ( {\cal L}_{\xi} A_{IJ}) =
\delta_i A_{IJ} + {\cal L}_{\xi} (\delta_i A_{IJ}) \, ,\nonumber\\
\delta_i {B'}^{IJ} & = & \delta_i B^{IJ}+ \delta_i ({\cal L}_{\xi} B^{IJ}) =
\delta_i B^{IJ}+ {\cal L}_{\xi}(\delta_i B^{IJ}) \, .
\end{eqnarray}
Therefore, at first order in the gauge parameters,
\begin{eqnarray}
{\omega'} & := &  \left ( \delta_1 {B'}^{IJ} \wedge \delta_2 {A'}_{IJ} -
\delta_2
{B'}^{IJ} \wedge \delta_1 {A'}_{IJ} \right ) \nonumber\\
& = & \omega + {\cal L}_{\xi} \omega \, .
\end{eqnarray}
Finally,
\begin{eqnarray}
{\Omega'}_{\Sigma} & := & \int_{\Sigma} \omega' =  \int_{\Sigma} \omega +
\int_{\Sigma} \left ( \xi \cdot d \omega + d (\xi \cdot \omega) \right )=
\Omega_{\Sigma} + \int_{\Sigma} \left ( \xi \cdot d \omega + d (\xi \cdot
\omega) \right ) \, .
\end{eqnarray}
If the linearized Eulerian derivatives vanish, i.e., if $\delta_i (F_{IJ})=0$
and $\delta_i (D B^{IJ}) =0$, $i=1,2$ then $d \omega=0$, and so
\begin{eqnarray}
{\Omega'}_{\Sigma} & = & \Omega_{\Sigma} + \int_{\Sigma} d (\xi \cdot \omega)
= \Omega_{\Sigma} + \int_{\partial \Sigma} \xi \cdot \omega \, .
\end{eqnarray}
Note that $A_{IJ}$ and $B^{IJ}$ in $\delta_i (F_{IJ})=0$ and $\delta_i (D
B^{IJ}) =0$ are {\it not} required to be solutions to the equations of motion
$F_{IJ}=0$ and $D B^{IJ}=0$, i.e., $\delta_i A_{IJ}$ and $\delta_i B^{IJ}$ are
tangent to ${\mathscr{F}}$ but not to $\overline{\mathscr{F}}$. Thus, if
\begin{eqnarray}
\int_{\partial \Sigma} \xi \cdot \omega =0 \, , \label{condi}
\end{eqnarray}
then
\begin{eqnarray}
{\Omega'}_{\Sigma} = \Omega_{\Sigma} \, .
\end{eqnarray}
This result can be summarized in the following:

{\it Proposition}: If the linearized Eulerian derivatives $\delta_i
(F_{IJ})=0$ and $\delta_i (D B^{IJ}) =0$ hold then Eq. (\ref{condi}) is a
necessary and sufficient condition for $\Omega_{\Sigma}$ be invariant under
the transformation associated with infinitesimal diffeomorphisms, i.e., if the
linearized Eulerian derivatives $\delta_i (F_{IJ})=0$ and $\delta_i (D B^{IJ})
=0$ hold then Eq. (\ref{condi}) is a necessary and sufficient condition for
infinitesimal diffeomorphisms to be canonical transformations.

It is clear that in the particular case when $\Sigma$ has no boundary, i.e.,
$\partial \Sigma= \emptyset$ then Eq. (\ref{condi}) holds without any
additional restrictions on $\xi$.

{\it Degenerate directions}: The starting point is the expression for the
presymplectic $3$-form with $\delta_1 =\delta$ an arbitrary variation and
$\delta_2$ is taken as the variation induced by the Lie derivative on the
dynamical fields. From Eqs. (\ref{PreStru}) and (\ref{4diff}),
\begin{eqnarray}
\omega (\delta A ,\delta B, {\cal L}_{\xi} A , {\cal L}_{\xi} B) & = & \left (
\delta B^{IJ} \wedge {\cal L}_{\xi} A_{IJ} - {\cal L}_{\xi} (B^{IJ} \wedge
\delta A_{IJ}) + B^{IJ} \wedge \delta ( {\cal L}_{\xi}  A_{IJ}) \right ) \, .
\label{Wald1}
\end{eqnarray}
Now, by taking $\delta ={\cal L}_{\xi}$ in the expression for the
presymplectic current potential $3$-form of Eq. (\ref{PrePotBF}) and computing
its variation one has
\begin{eqnarray}
\delta {\bf \Theta} (B, {\cal L}_{\xi}A) & = &  \delta B^{IJ} \wedge {\cal
L}_{\xi} A_{IJ} +  B^{IJ} \wedge \delta ({\cal L}_{\xi} A_{IJ}) \, .
\label{Wald2}
\end{eqnarray}
Inserting the right-hand side of Eq. (\ref{Wald2}) into the right-hand side of
Eq. (\ref{Wald1}) one gets
\begin{eqnarray}
\omega (\delta A ,\delta B, {\cal L}_{\xi} A , {\cal L}_{\xi} B) & = & \delta
{\bf \Theta} (B, {\cal L}_{\xi}A) - {\cal L}_{\xi} ( B^{IJ} \wedge \delta
A_{IJ}) \nonumber\\
& = & \delta {\bf \Theta} (B, {\cal L}_{\xi}A) - {\cal L}_{\xi} {\bf \Theta}
(B,\delta A) \, . \label{Wald3}
\end{eqnarray}
On the other hand, the variation of the Noether current $3$-form is
\begin{eqnarray}
\delta {\bf J}_N [A,B,\xi] & = & \delta {\bf \Theta}(B,{\cal L}_{\xi} A) - \xi
\cdot \delta {\bf L}
\nonumber\\
& = & \delta {\bf \Theta}(B,{\cal L}_{\xi} A) - \xi \cdot \left( ( \delta
B^{IJ}) \wedge F_{IJ} -  \left ( D B^{IJ} \right ) \wedge \delta A_{IJ} + d
{\bf \Theta} (B, \delta A) \right ) \nonumber\\
& = & \omega (\delta A ,\delta B, {\cal L}_{\xi} A , {\cal L}_{\xi} B) + {\cal
L}_{\xi} {\bf \Theta} (B,\delta A) \nonumber\\
& & \mbox{} - \xi \cdot \left( ( \delta B^{IJ}) \wedge F_{IJ} -  \left ( D
B^{IJ} \right ) \wedge \delta A_{IJ} \right )
- \xi \cdot d {\bf \Theta} (B, \delta A) \nonumber\\
& = & \omega (\delta A ,\delta B, {\cal L}_{\xi} A , {\cal L}_{\xi} B) - \xi
\cdot \left( ( \delta B^{IJ}) \wedge F_{IJ} -  \left ( D B^{IJ} \right )
\wedge \delta A_{IJ} \right ) \nonumber\\
& & \mbox{} + d \left ( \xi \cdot {\bf \Theta}(B, \delta A) \right ) \, .
\label{Wald4}
\end{eqnarray}
To get the second line on the right-hand side, Eq. (\ref{VarBFLag}) was used
while Eq. (\ref{Wald3}) was used to get the third line. Inserting the explicit
expression for $\delta {\bf J}_N [A,B,\xi]$ given in Eq. (\ref{VarJ}) into the
left-hand side of Eq. (\ref{Wald4}) one has
\begin{eqnarray}
\omega (\delta A ,\delta B, {\cal L}_{\xi} A , {\cal L}_{\xi} B) & = & d \left
( \delta {\bf Q}_N [A,B,\xi] - \xi \cdot {\bf \Theta}(B, \delta A) \right ) -
(\xi \cdot A_{IJ}) \delta DB^{IJ} \nonumber\\
& & \mbox{} -  (\xi \cdot B^{IJ} ) \wedge \delta F_{IJ} + (\delta B^{IJ})
\wedge (\xi \cdot F_{IJ}) \nonumber\\
& & \mbox{} -  (\xi \cdot D B^{IJ}) \wedge \delta A_{IJ} \, .
\end{eqnarray}
Note that, in contrast to Eqs. (\ref{Gauomega}) and (\ref{chiomega}), in the
case of diffeomorphisms the symplectic inner product between $\delta_{\xi}$
and an arbitrary variation $\delta$ involves both Eulerian derivatives and the
linearized Eulerian derivatives. One has the following:

{\it Proposition}: Let $(A_{IJ}, B^{IJ})$ be a point in
$\overline{\mathscr{F}}$; let $(\delta A_{IJ}, \delta B^{IJ})$ be a solution
to the linearized Eulerian derivatives at $(A_{IJ}, B^{IJ})$ [i.e., $(\delta
A_{IJ}, \delta B^{IJ})$ are such that $\delta(D B^{IJ})=0$ and $\delta
F_{IJ}=0$ and are tangent to $\overline{\mathscr{F}}$ at $(A_{IJ}, B^{IJ})$].
Then, we have
\begin{eqnarray}
\omega (\delta A ,\delta B, {\cal L}_{\xi} A , {\cal L}_{\xi} B) & = & d \left
( \delta {\bf Q}_N [A,B,\xi] - \xi \cdot {\bf \Theta}(B, \delta A) \right ) \,
,
\end{eqnarray}
and thus, integrating on $\Sigma$,
\begin{eqnarray}
\Omega_{\Sigma}(\delta A ,\delta B, {\cal L}_{\xi} A, {\cal L}_{\xi} B) &:= &
\int_{\Sigma} \omega (\delta A ,\delta B, {\cal L}_{\xi}
A, {\cal L}_{\xi} B) \nonumber\\
& = & \oint_{\partial \Sigma} \left ( \delta {\bf Q}_N [A,B,\xi] - \xi \cdot
{\bf \Theta}(B, \delta A) \right ) \, .
\end{eqnarray}
Inserting the explicit expressions for ${\bf Q}_N[A,B,\xi]$ and ${\bf
\Theta}(B,\delta A)$ one has
\begin{eqnarray}
\Omega_{\Sigma} (\delta A , \delta B, {\cal L}_{\xi} A, {\cal L}_{\xi} B) & =
& \oint_{\partial \Sigma} \left [ \delta (  (\xi \cdot A_{IJ}) B^{IJ}) - \xi
\cdot ( B^{IJ} \wedge \delta A_{IJ}) \right
]\nonumber\\
& = & \oint_{\partial \Sigma} \left [  (\xi \cdot A_{IJ}) \delta B^{IJ} -
(\xi \cdot B^{IJ}) \wedge \delta A_{IJ}  \right ] \, . \label{KeyDiff}
\end{eqnarray}
Some remarks follow: (1) first of all, Eq. (\ref{KeyDiff}) tells us that, in
the context of the covariant canonical formalism, not all diffeomorphisms are
to be regarded as gauge because the right-hand side of (\ref{KeyDiff}) will
{\it not} vanish for any $\xi$, (2) note that if $(A_{IJ}, B^{IJ}) \in
\overline{\mathscr{F}}$ then $\xi \cdot A_{IJ} \neq 0$ and $\xi \cdot
B^{IJ}\neq 0$ in the generic case. Moreover, note that the right-hand side of
(\ref{KeyDiff}) vanishes for all $(A_{IJ},B^{IJ})$ of $\overline\mathscr{F}$
and for all tangent variation $(\delta A_{IJ}, \delta B^{IJ} )$ to
$\overline{\mathscr{F}}$ in $(A_{IJ},B^{IJ})$ if and only if $\xi$ vanishes at
the boundary $\partial\Sigma$, $\xi\mid_{\partial \Sigma}=0$. Thus, just those
diffeomorphisms which are the identity at $\partial \Sigma$ must be regarded
as gauge. One could say that the gauge transformation is broken at $\partial
\Sigma$ in the sense that $\xi\mid_{\partial \Sigma}=0$. However, from that
perspective one would be {\it a priori} assuming  that all diffeomorphisms are
gauge which as the previous analysis shows is not the case. Let $\zeta$ be a
diffeomorphism such that it does not vanish at $\partial \Sigma$. The full set
of these $\zeta$'s span the {\it boundary symmetry group}. Thus, the covariant
canonical formalism tells us that the {\it boundary symmetry group} is {\it
not} a gauge group (see also Ref. \cite{ABR} to understand the role of
diffeomorphisms in the case of general relativity).

{\it  Existence of a Hamiltonian}: For variations $\delta A_{IJ}$ with compact
support in the interior of $\Sigma$, i.e., $\delta A_{IJ}\mid_{\partial
\Sigma}=0$,
\begin{eqnarray}
\Omega_{\Sigma} = \delta \oint_{\partial \Sigma}  (\zeta \cdot A_{IJ}) B^{IJ}
= \delta \oint_{\partial\Sigma} {\bf Q}[A,B,\zeta] \, ,
\end{eqnarray}
which means that a Hamiltonian conjugate to $\zeta$ on $\Sigma$ exists and
that its Hamiltonian density is precisely the Noether potential ${\bf
Q}[A,B,\zeta]$. We have assumed that $\zeta$ does {\it not} vanish at
$\partial \Sigma$ and thus, by definition, $\zeta$ is not a degenerate
direction.

On the other hand, for variations $\delta B_{IJ}$ with compact support in the
interior of $\Sigma$, i.e., $\delta B^{IJ}\mid_{\partial \Sigma}=0$,
\begin{eqnarray}
\Omega_{\Sigma} & = & \delta \oint_{\partial \Sigma} -  (\zeta \cdot B^{IJ})
\wedge A_{IJ}\, ,
\end{eqnarray}
so there exists a Hamiltonian conjugate to $\zeta$ on $\Sigma$.

{\it Relationship between the Noether currents}: It is possible to compare the
Noether current associated with diffeomorphisms, ${\bf J}_N [A,B,\xi]$, with
the currents associated to the fundamental set of gauge symmetries
\begin{eqnarray}
{\bf J}_N [A,B,\xi] & = & B^{IJ} \wedge {\cal L}_{\xi} A_{IJ} - \xi \cdot (
B^{IJ} \wedge F_{IJ} ) \nonumber\\
& = &  B^{IJ} \wedge \left [ \xi \cdot F_{IJ} + D \varepsilon_{IJ} \right ] -
 (\xi \cdot B^{IJ}) \wedge F_{IJ} -  B^{IJ} \wedge (\xi \cdot F_{IJ})
\nonumber\\
& = &  B^{IJ} \wedge D \varepsilon_{IJ} -  \chi^{IJ} \wedge F_{IJ}
\nonumber\\
& = & {\bf J}_N [A,B,\varepsilon] - {\bf J}_N [A,\chi] \, .
\end{eqnarray}
In a more appropriate notation
\begin{eqnarray}
{\bf J}_N [A,B,\xi ] = \left ( {\bf J}_N [A,B,\varepsilon] - {\bf J}_N
[A,\chi] \right ) \mid_{\varepsilon_{IJ} = \xi \cdot A_{IJ}, \chi^{IJ} = \xi
\cdot B^{IJ}} \, .
\end{eqnarray}
Note also that ${\bf J}_N [A,B,\xi ]= {\bf J}_N [A,B,\varepsilon]$ because
${\bf J}_N [A,\chi]=0$ on-shell (i.e., if $F_{IJ}=0$). Moreover, note that
\begin{eqnarray}
{\bf Q} [A,B,\xi] & = &  (\xi \cdot A_{IJ}) B^{IJ} =  \varepsilon_{IJ} B^{IJ}
= {\bf Q} [B,\varepsilon] \, .
\end{eqnarray}

\section{BF theory plus a cosmological constant}
The four-dimensional BF theory with $SO(3,1)$ as the internal group and
supplemented with a cosmological constant $\Lambda$ is defined by the
equations of motion
\begin{eqnarray}\label{esol}
F_{IJ}= 2 \Lambda \ast B_{IJ} \, , \quad D B_{IJ}=0 \, ,
\end{eqnarray}
where $\ast B_{IJ} =\frac{1}{2}\epsilon_{IJKL} B^{KL}$ is the dual of
$B^{IJ}$. If $SO(4)$ were taken as the internal group then $\eta_{IJ}
\longrightarrow \delta_{IJ}$, the connection were valued in the Lie algebra of
$SO(4)$ and $\ast^2=+1$. Equations (\ref{esol}) can be obtained from the
action principle \cite{Gary,DePietri}
\begin{eqnarray}
S[A, B] & = &  \int_{{\cal M}} B^{IJ} \wedge F_{IJ} [A] - \Lambda \int_{\cal
M} B^{IJ} \wedge \ast B_{IJ} \, . \label{BF+CC}
\end{eqnarray}
Thus, in contrast to BF theory, the space of solutions to the equations of
motion $\overline{\mathscr{F}}_{BF+\Lambda}$ is now defined by Eq.
(\ref{esol}). To get the geometry, one needs to compute the first order
variation of the Lagrangian 4-form ${\bf L}[A,B]= B^{IJ} \wedge F_{IJ} [A] -
\Lambda B^{IJ} \wedge \ast B_{IJ}$,
\begin{eqnarray}
\delta {\bf L}[A,B] & = & (\delta B^{IJ}) \wedge \left ( F_{IJ} - 2 \Lambda
\ast B_{IJ} \right ) - \left ( D B^{IJ} \right ) \wedge \delta A_{IJ} + d {\bf
\Theta} (B, \delta A) \, , \label{BFcc}
\end{eqnarray}
from which the {\it presymplectic potential} $3$-form
\begin{eqnarray}
{\bf \Theta} (B,\delta A) := B^{IJ} \wedge \delta A_{IJ} \, ,
\label{PrePotBFcc}
\end{eqnarray}
is read off. Therefore, the presymplectic 3-form $\omega$ is the same of the
BF theory. The symplectic structure induced on
$\overline{\mathscr{F}}_{BF+\Lambda}$ is simply the pullback to
$\overline{\mathscr{F}}_{BF+\Lambda}$  of the curl of ${\bf \Theta}$ on the
kinematical phase space.

{\it Degenerate directions}:
\begin{enumerate}
\item
The symplectic inner product between $\delta_1 =\delta$ and
$\delta_2=\delta_{\varepsilon}$, $\omega(\delta A,\delta B,
\delta_{\varepsilon} A, \delta_{\varepsilon} B)$, has the same analytical form
of the BF theory.
\item
The symplectic inner product between $\delta_1 =\delta$ and $\delta_2
=\delta_{\chi}$ where $\delta_{\chi} A_{IJ} = \Lambda \varepsilon_{IJKL}
\chi^{KL} $ and $\delta_{\chi} B^{IJ}= D \chi^{IJ}$ (Ref. \cite{MonCQG03}) is
now
\begin{eqnarray} \omega(\delta A,\delta B, \delta_{\chi} A, \delta_{\chi} B)
& = &  d (\delta \left ( - \chi^{IJ} \wedge  A_{IJ} \right )) -  \chi^{IJ}
\wedge \delta \left ( F_{IJ} - 2 \Lambda \ast B_{IJ} \right )
\label{chiomegacc} \, .
\end{eqnarray}
\item
The Noether current associated with diffeomorphisms acquires the form
\cite{MonCQG03}
\begin{eqnarray}
{\bf J}_N [A,B,\xi] & = & d {\bf Q}_N [A,B,\xi] - (\xi \cdot A_{IJ})  D B^{IJ}
- (\xi \cdot B^{IJ}) \wedge \left ( F_{IJ} - 2 \Lambda \ast B_{IJ} \right ) \,
, \label{NCDiffcc}
\end{eqnarray}
with the Noether current potential 2-form ${\bf Q}_N [A,B,\xi]$ having the
same analytical form than the one of the BF theory. Due to the fact that
$\omega (\delta A ,\delta B, {\cal L}_{\xi} A , {\cal L}_{\xi} B)  =  d \left
( \delta {\bf Q}_N [A,B,\xi] - \xi \cdot {\bf \Theta}(B, \delta A) \right )$
if the equations of motion and the linearized Eulerian derivatives hold and
because ${\bf Q}_N [A,B,\xi]$ and ${\bf \Theta}(B, \delta A)$ have the same
form of the BF theory, then analysis of the degenerate directions is the same
of the BF theory. Finally, note that now $\delta_{\xi}$,
$\delta_{\varepsilon}$, and $\delta_{\chi}$ are not related by (\ref{mau}) but
by $\delta_{\xi} A_{IJ} = \xi \cdot \left ( F_{IJ} - 2 \Lambda \ast B_{IJ}
\right ) + \delta_{\varepsilon} A_{IJ} +\delta_{\chi} A_{IJ}$ while
$\delta_{\xi} B^{IJ} = \xi \cdot D B^{IJ} + \delta_{\varepsilon} B^{IJ} +
\delta_{\chi} B^{IJ}$ retains his form with $\varepsilon_{IJ}= \xi \cdot
A_{IJ}$, $\chi^{IJ}= \xi \cdot B^{IJ}$.
\end{enumerate}

\section{BF theory plus the second Chern character}
Now, the action we consider is the action for BF theory supplemented with the
second Chern character
\begin{eqnarray}
S[A, B] & = &  \int_{{\cal M}} B^{IJ} \wedge F_{IJ} [A] + \theta \int_{{\cal
M}} F^{IJ} [A] \wedge F_{IJ} [A]  \, , \label{BFChern}
\end{eqnarray}
with $\theta$ a parameter. The first order variation of the Lagrangian 4-form
${\bf L}[A,B] = B^{IJ} \wedge F_{IJ} + \theta F^{IJ} \wedge F_{IJ}$ is
\begin{eqnarray}\label{VarBFChern}
\delta {\bf L} [A,B] = \delta B^{IJ} \wedge F_{IJ} - \left (  D B^{IJ} + 2
\theta D F^{IJ} \right ) \wedge \delta A_{IJ} + d {\bf \Theta} \, ,
\end{eqnarray}
where
\begin{eqnarray}
{\bf \Theta} = \left (  B^{IJ} + 2 \theta F^{IJ} \right ) \wedge \delta A_{IJ}
\, ,
\end{eqnarray}
is the presymplectic potential 3-form [cf. Eq. (\ref{PrePotBF})]. From Eq.
(\ref{VarBFChern}), it is clear that the equations of motion $F_{IJ}=0$ and $
D B^{IJ} + 2 \theta D F^{IJ}=0$ coming from Eq. (\ref{BFChern}) reduce to
those of the BF theory because the Bianchi identities $D F_{IJ}=0$ always
hold. Thus, the inclusion of the second Chern character does not modify the
classical dynamics of the BF theory, as expected. This means that the space of
solutions to the equations of motion $\overline{\mathscr{F}}$ is the same for
both theories. Nevertheless, in spite of the fact that the equations of motion
are the same, the presymplectic 3-form changes. For the present case one has
\begin{eqnarray}\label{clave}
\omega & = & \delta_1 \left ( B^{IJ} + 2 \theta F^{IJ} \right ) \wedge
\delta_2 A_{IJ} - \delta_2 \left ( B^{IJ} + 2 \theta F^{IJ} \right ) \wedge
\delta_1 A_{IJ}
\end{eqnarray}
[cf. Eq. (\ref{PreStru})]. Therefore, the presymplectic 3-forms coming from
Eqs. (\ref{BFaction}) and (\ref{BFChern}) are distinct. On the space of
solutions $\overline{\mathscr{F}}$, the symplectic structure of Eq.
(\ref{clave}) is the same as the symplectic structure of Eq. (\ref{PreStru}),
of course (see also Ref. \cite{me}).

{\it Degenerate directions}:
\begin{enumerate}
\item
The symplectic inner product between $\delta_1=\delta$ and
$\delta_2=\delta_{\varepsilon}$ is now
\begin{eqnarray}
\omega(\delta A, \delta B , \delta_{\varepsilon} A , \delta_{\varepsilon} B) &
= & d \left (  \,\, \varepsilon_{IJ} \delta B^{IJ} + 2 \theta \,\, \delta
A^{IJ} \wedge D \varepsilon_{IJ} \right ) -  \varepsilon_{IJ} \delta (D
B^{IJ})
\end{eqnarray}
and so
\begin{eqnarray}\label{sipChernI}
\Omega_{\Sigma} (\delta A, \delta B , \delta_{\varepsilon} A ,
\delta_{\varepsilon} B) & = & -  \int_{\Sigma} \varepsilon_{IJ} \delta D
B^{IJ} + \int_{\partial \Sigma} \left ( \,\,\varepsilon_{IJ} \delta B^{IJ} + 2
\theta \,\, \delta A^{IJ} \wedge D \varepsilon_{IJ} \right ) \, .
\end{eqnarray}
Therefore, if the linearized Eulerian derivative $\delta (D B^{IJ})$ vanishes,
$\delta (D B^{IJ})=0$, and the arbitrary variations $(\delta A^{IJ}, \delta B^
{IJ})$ have compact support in the interior of $\Sigma$, $\delta A^{IJ}
\mid_{\partial \Sigma}=0$ and $\delta B^{IJ} \mid_{\partial\Sigma}=0$, then
\begin{eqnarray}\label{kerChernI}
\Omega_{\Sigma} (\delta A,\delta B, \delta_{\varepsilon} A,
\delta_{\varepsilon} B) & = & 0 \, ,
\end{eqnarray}
without imposing any additional restrictions on the gauge parameters
$\varepsilon_{IJ}$. The integral over $\partial \Sigma$ in Eq.
(\ref{sipChernI}) also vanishes if both the gauge parameters
$\varepsilon_{IJ}$ vanish and satisfy $D \varepsilon_{IJ}=0 $ at
$\partial\Sigma$.
\item
The symplectic inner product between $\delta_1=\delta$ and $\delta_2
=\delta_{\chi}$ is the same as the one of the BF theory.
\item
Now, we consider diffeomorphisms. The Noether current associated with
diffeomorphisms acquires the form
\begin{eqnarray}
{\bf J}_N [A,B,\xi] & = & d {\bf Q}_N [A,B,\xi] - (\xi \cdot A_{IJ}) \left ( D
B^{IJ}+ 2\theta D F^{IJ} \right ) -  (\xi \cdot B^{IJ}) \wedge F_{IJ} \, ,
\label{NCChern}
\end{eqnarray}
where
\begin{eqnarray}
{\bf Q}_N [A,B,\xi] = (\xi \cdot A_{IJ}) \left (  B^{IJ} + 2 \theta F^{IJ}
\right ) \, ,
\end{eqnarray}
is the {\it Noether current potential} $2$-form. Therefore, the Noether charge
is the same as in the BF theory if the equations of motion are satisfied.
Moreover, if both the equations of motion and the linearized equations of
motion hold, then symplectic inner product between $\delta_1 =\delta$ and
$\delta_2=\delta_{\xi}$ becomes
\begin{eqnarray}
\Omega_{\Sigma}(\delta A ,\delta B, {\cal L}_{\xi} A, {\cal L}_{\xi} B) & = &
\oint_{\partial \Sigma} \left ( \delta {\bf Q}_N [A,B,\xi] - \xi \cdot {\bf
\Theta}(B, \delta A) \right ) \, .
\end{eqnarray}
Inserting the explicit expressions for ${\bf Q}_N[A,B,\xi]$ and ${\bf
\Theta}(B,\delta A)$ one has
\begin{eqnarray}
\Omega_{\Sigma} (\delta A , \delta B, {\cal L}_{\xi} A, {\cal L}_{\xi} B) & =
& \oint_{\partial \Sigma} \left [ \delta \left ( \xi \cdot A_{IJ} \left (
B^{IJ} + 2\theta F^{IJ} \right ) \right ) \right. \nonumber\\
& & \mbox{} \left.  - \xi \cdot \left ( \left ( B^{IJ}
+ 2\theta F^{IJ} \right ) \wedge \delta A_{IJ} \right ) \right ]\nonumber\\
& = & \oint_{\partial \Sigma} \left [  (\xi \cdot A_{IJ}) \delta B^{IJ} -
(\xi \cdot B^{IJ}) \wedge \delta A_{IJ}  \right ] \, ,
\end{eqnarray}
because $F^{IJ}=0$ and $\delta F^{IJ}=0$, by hypothesis. Last equation has the
same analytical form as the one of the BF theory. Therefore, the inclusion of
the second Chern character does not modify the degenerate directions in the
case of diffeomorphisms.
\end{enumerate}

\section{Conclusions and perspectives}
To conclude, we emphasize the role that the Bianchi identities $DF_{IJ}=0$
play in four-dimesional BF theories. On the one hand, they are the cause of
having the symmetry $\delta_{\chi}$ in the various four-dimensional BF
theories already discussed. On the other hand, in the case of the BF theory
with a nonvanishing cosmological constant $\Lambda$ the combination of the
Bianchi identities together with the equation of motion $F_{IJ}= 2 \Lambda
\ast B_{IJ}$ ``generates" dynamics for the $B^{IJ}$ fields in the sense that
they imply $D B_{IJ}=0$. This fact is the origin of the reducibility of the
corresponding first class constraints of the theory in Dirac's canonical
analysis. This same phenomenon appears, in essence, in the action
\cite{MonCQG03}
\begin{eqnarray}\label{mmm}
S[A,B,\phi] = \int_{\cal M} B^{IJ} \wedge F_{IJ} - \frac12 \phi_{IJKL} B^{IJ}
\wedge B^{KL}\, ,
\end{eqnarray}
with $\phi_{IJKL}=-\phi_{JIKL}=-\phi_{IJLK}=\phi_{KLIJ}$ where the combination
of the Bianchi identities and the equations of motion $F_{IJ}=\phi_{IJKL}
B^{KL}$ and $D B^{IJ}=0$ generates dynamics for the $\phi_{IJKL}$ fields in
the sense that these equations imply $(D \phi_{IJKL}) \wedge B^{KL}=0$. From
the lesson learned from the case of the BF theory plus a cosmological constant
$\Lambda$, we would expect that the theory defined by Eq. (\ref{mmm}) has also
reducibility in the constraints in the context of Dirac's analysis. We
consider the present analysis as a first step towards the covariant canonical
analysis of BF gravity \cite{Capo}.

\section*{Acknowledgements}
Warm thanks to J.D. Vergara for discussions about reducibility of constraints
and to R. Cartas-Fuentevilla for discussions about the covariant canonical
formalism. This work was supported in part by CONACyT grant no.
SEP-2003-C02-43939.

\appendix

\section{Review of Dirac's canonical analysis for four-dimensional BF theory}
To compare some results of this Appendix with some results of the covariant
canonical formalism one must make the changes $\varepsilon_{IJ}\longrightarrow
- \varepsilon_{IJ}$ and $\varepsilon^{IJ}_a \longrightarrow - \chi^{IJ}_a$ in
this Appendix.

(1) {\it BF theory}: By making the 3+1 decomposition, a straightforward
computation shows that the action (\ref{BFaction}) acquires the form
\begin{eqnarray}
S [A_{aIJ}, {\widetilde \Pi}^{aIJ} , \lambda_{IJ} , \lambda_a\,^{IJ} ] & = &
\int_{\cal M} d ^4 x \left [ {\dot A}_{aIJ} {\widetilde \Pi}^{aIJ} -
\lambda_{IJ} D_a {\widetilde \Pi}^{aIJ} - {\lambda}_a\,^{IJ} \left ( \frac12
{\widetilde \eta}^{abc} F_{bcIJ} \right ) \right ] \nonumber\\
&& \mbox{} + \int_{\cal M} d ^4 x \,
\partial_a ( \lambda_{IJ} {\widetilde \Pi}^{aIJ}) \, , \label{HamBF}
\end{eqnarray}
where the phase space variables $(A_{aIJ} , {\widetilde \Pi}^{bKL})$ and
Lagrange multipliers $\lambda_{IJ}$ and $\lambda_a\,^{IJ}$ are defined in
terms of the initial configuration variables as ${\widetilde \Pi}^{aIJ}:=
\frac12 {\widetilde \eta}^{abc} B_{bc}\,^{IJ}$, $ \lambda_{IJ}:= - A_{0IJ}$,
${\lambda}_a\,^{IJ}:=- B_{0a}\,^{IJ}$, $D_a {\widetilde \Pi}^{aIJ}:=
\partial_a {\widetilde \Pi}^{aIJ} + A_{a}\,^I\,_K {\widetilde \Pi}^{aKJ} +
A_a\,^J\,_K {\widetilde \Pi}^{aIK}$. If the spacetime ${\cal M}$ has the
topology ${\cal M}=\Sigma \times R$ and $\Sigma$ has no boundary the second
integral on the right-hand side of (\ref{HamBF}) can be neglected. The
lower-case letters $a,b$ are space ones and run from $1$ to $3$. Notice that
the Lorentz indices $I,J$ are not split holding in this way the full Lorentz
group. The variation of Eq. (\ref{HamBF}) with respect to the phase space
variables yields the equations of motion
\begin{eqnarray}
{\dot A}_{aIJ} & = & -D_a \lambda_{IJ} \, , \nonumber\\
{\dot {\widetilde \Pi}}^{aIJ} & = & 2 \lambda^{[I}\,_K {\widetilde \Pi}^{a\,
KJ]} - {\widetilde \eta}^{abc} D_b \lambda_c\,^{IJ}\, ,
\end{eqnarray}
with $D_a \lambda_{IJ}=\partial_a \lambda_{IJ} - A_a\,^K\,_I\lambda_{KJ}-
A_{a}\,^K\,_J \lambda_{IK} $. The variation with respect to the Lagrange
multipliers gives the constraints
\begin{eqnarray}
{\widetilde \Psi}^{IJ} & := & D_a {\widetilde \Pi}^{aIJ} \approx 0\, , \quad
{\widetilde \Psi}^a\,_{IJ} :=  \frac12 {\widetilde \eta}^{abc} F_{bcIJ} (A)
\approx 0 \, , \label{ConstBF}
\end{eqnarray}
which are first class. The infinitesimal gauge transformation generated by the
Gauss constraint ${\widetilde \Psi}^{IJ}$ is
\begin{eqnarray}
{A'}_{aIJ} & = & A_{aIJ}- D_a \varepsilon_{IJ} \, , \quad {\widetilde
{\Pi'}}^{aIJ} =  {\widetilde \Pi}^{aIJ} + \varepsilon^{IM} {\widetilde
\Pi}^a\,_M\,^{J} - \varepsilon^{JM} {\widetilde \Pi}^{a}\,_M\,^I \, ,
\end{eqnarray}
and
\begin{eqnarray}
{A'}_{aIJ} & = & A_{aIJ} \, , \quad {\widetilde {\Pi'}}^{aIJ}  =  {\widetilde
\Pi}^{aIJ} - {\widetilde \eta}^{abc} D_b \varepsilon^{IJ}\,_c\, ,
\end{eqnarray}
is the gauge transformation generated by the constraint ${\widetilde
\Psi}^a\,_{IJ}$. However, even though the constraints ${\widetilde \Psi}^{IJ}$
are irreducible the constraints ${\widetilde \Psi}^a\,_{IJ}$ are not, i.e.,
they are {\it reducible}. This is so because the Bianchi identities $D
F_{IJ}=0$ imply the relationship among the ${\widetilde \Psi}^a\,_{IJ}$'s,
\begin{eqnarray}
{\widetilde \Phi}_{IJ} & := & D_a {\widetilde \Psi}^a\,_{IJ} = 0\, .
\end{eqnarray}
The counting of physical degrees of freedom is as follows. There are $3\times
6=18$ configuration variables $A_{aIJ}$ and $6 + [(3\times 6)-6] =18$ {\it
independent} first class constraints. Therefore, the system has no local
degrees of freedom \cite{Caicedo}. Alternatively, the independent number of
gauge parameters is $18 =6$ (the $\varepsilon_{IJ}$'s) $+$ $12$ ($=18-6$
independent gauge parameters from $\varepsilon^{IJ}_a$).

(2) {\it BF theory plus a cosmological constant}: By performing the $3+1$
decomposition the action (\ref{BF+CC}) can be written as
\begin{eqnarray}
S [A_{aIJ}, {\widetilde \Pi}^{aIJ} , \lambda_{IJ} , \lambda_a\,^{IJ} ] & = &
\int d ^4 x \left [ {\dot A}_{aIJ} {\widetilde \Pi}^{aIJ} - \lambda_{IJ} D_a
{\widetilde \Pi}^{aIJ}  \right. \nonumber\\
& & \mbox{} \left. - \lambda_a\,^{IJ} \left ( \frac12 {\widetilde \eta}^{abc}
F_{bcIJ} - \Lambda \varepsilon_{IJKL} {\widetilde \Pi}^{aKL} \right )\right
]\, . \label{Yang}
\end{eqnarray}
The equations of motion are
\begin{eqnarray}
{\dot A}_{aIJ} & = & - D_a \lambda_{IJ} - \Lambda \varepsilon_{IJKL}
\lambda_a\,^{KL} \, ,
\nonumber\\
{\dot {\widetilde \Pi}}^{aIJ} & = & 2 \lambda^{[I}\,_K {\widetilde \Pi}^{a\,
KJ]} - {\widetilde \eta}^{abc} D_b \lambda_c\,^{IJ}\, .
\end{eqnarray}
The constraints are
\begin{eqnarray}
{\widetilde \Psi}^{IJ} & := & D_a {\widetilde \Pi}^{aIJ}\, , \quad {\widetilde
\Psi}^a\,_{IJ} :=  \frac12 {\widetilde \eta}^{abc} F_{bcIJ} - \Lambda
\varepsilon_{IJKL} {\widetilde \Pi}^{aKL} \, .
\end{eqnarray}
The evolution of the constraints provides no more constraints. To compute the
algebra of constraints it is convenient to smear them
\begin{eqnarray}
\Psi[u] & := & \int d^3 x u_{IJ} {\widetilde \Psi}^{IJ} \, , \quad \Psi [N] :=
\int d^3 x N^{IJ}\,_a {\widetilde \Psi}^a\,_{IJ} \, .
\end{eqnarray}
The constraint algebra is
\begin{eqnarray}
\{ \Psi[u] , \Psi [v] \} & = & \Psi [[u,v]]\, , \quad \{ \Psi[u] , \Psi[N] \}
=  \Psi [[u,N]]  \, , \quad \{ \Psi[N] , \Psi[M] \}  =  0\, ,
\end{eqnarray}
with $[u,v]_{IJ}:= u_I\,^M v_{MJ}- u_J\,^M v_{MI}$, $[u,N]^{IJ}\,_a= u^I\,_K
N^{KJ}\,_a - u^J\,_K N^{KI}\,_a$. The infinitesimal gauge transformation
generated by the Gauss constraint ${\widetilde \Psi}^{IJ}$ is
\begin{eqnarray}
{A'}_{aIJ} & = & A_{aIJ}- D_a \varepsilon_{IJ} \, , \quad {\widetilde
{\Pi'}}^{aIJ} =  {\widetilde \Pi}^{aIJ} + \varepsilon^{IM} {\widetilde
\Pi}^a\,_M\,^{J} - \varepsilon^{JM} {\widetilde \Pi}^{a}\,_M\,^I \, ,
\end{eqnarray}
and
\begin{eqnarray}
{A'}_{aIJ} & = & A_{aIJ} - \Lambda \, \varepsilon_{IJKL} \,
\varepsilon^{KL}\,_a \, , \quad {\widetilde {\Pi'}}^{aIJ} =  {\widetilde
\Pi}^{aIJ} - {\widetilde \eta}^{abc} D_b \varepsilon^{IJ}\,_c\, ,
\end{eqnarray}
is the infinitesimal gauge transformation generated by the constraint
${\widetilde \Psi}^a\,_{IJ}$. Again, the Bianchi identities imply that the
constraints are reducible
\begin{eqnarray}
D_a {\widetilde \Psi}^a\,_{IJ} + \Lambda \varepsilon_{IJKL} {\widetilde
\Psi}^{KL} & = & 0\, . \label{RedEqII}
\end{eqnarray}
Like in pure BF gravity the system has $3\times 6 =18$ configuration variables
and $6+ [(3\times 6) - 6] =18$ independent first class constraints. Therefore,
the system has no local degrees of freedom, as expected because the addition
of a cosmological constant does not add local degrees of freedom. However, a
key difference with respect to the case without cosmological constant
$\Lambda$ is that there the constraints ${\widetilde \Psi}^a\,_{IJ}$ and
${\widetilde \Psi}^{IJ}$ are independent while in the present case they are
related through the reducibility equation given in Eq. (\ref{RedEqII}).
Moreover, due to the fact the reducibility equation involves now the Gauss
constraints too, there are $18$ independent gauge parameters among the $6$ of
$\varepsilon_{IJ}$'s and the $18$ of $\varepsilon^{IJ}_a$'s. One can take
these independent number of gauge parameters as the 18 of the
$\varepsilon^{IJ}_a$'s. By doing this, one might say that the local Lorentz
transformation is redundant if a cosmological constant is present.


\end{document}